# DESIGN AND CONSTRUCTION OF THE 3.2 MEV HIGH VOLTAGE COLUMN FOR DARHT II*

C. Peters, B. Elliott, S. Yu, S. Eylon, E. Henestroza,
Lawrence Berkeley National Laboratory, Berkeley, CA 94720

## 1 INTRODUCTION

A 3.2 MeV injector has been designed and built for the second axis of the Dual Axis Radiographic Hydrodynamic Test (Darht) facility at Los Alamos National Laboratory. Darht-II accelerator is a linac which produces a 2000 ampere electron pulse with a flattop width of at least 2-microseconds and emittance of less than 0.15 $\pi$ cm-rad normalized. The installation of the injector system is presently underway and commissioning is expected to begin in December 2000. The design and construction of the injector was completed by LBNL. The Marx Generator was designed and built under subcontract to LBNL. The installation of the insulating column will be occurring during August 2000.

The Darht-II injector High Voltage Column and Cathode Assembly are housed within a 4 meter diameter by 9 meter tall vacuum enclosure. The cathode assembly is supported on top of a vertically oriented insulating column. This arrangement is shown in figure 1. The volume outside of the column is vacuum and the inside volume is insulating oil. The current stalk passes up through the center of the column in coaxial fashion.

A novel method of construction has been used for the insulating column, which utilizes bonded mycalex insulating rings. This paper will describe the design, construction, and testing of the column. Mechanical aspects of the design will be emphasized.

## 2 REQUIREMENTS

The primary functions of the High voltage insulating column are to physically support the cathode assembly and to insulate the cathode from ground potential during the high voltage pulse. The cathode must be supported in a stable position to enable accurate alignment and stability with respect to the accelerator axis. The column must operate reliably electrically without significant breakdown along its length or radially to the vacuum vessel on the outside or to the current stalk on the inside. Electrical reliability of the column is crucial and any problems in this area will dramatically impact the operation of the Injector. The column provides a hermetic barrier between the insulating medium of the marx generator and the UHV volume of the cathode. Additionally, the column outgassing load must be consistent with the $5 \times 10^{-8}$ torr pressure requirement for the Injector.

## 3 DESIGN

The overall column design follows the conventional approach of using rings of insulating material separated by metal grading spacer rings. The triple points are shielded and resistors are used to grade the voltage on the spacer rings. Figure 2 shows the typical arrangement used throughout the column. Insulating rings are assembled into stacks. Nine stacks were made. Overall column length is 4.4 meters and diameter is 1.3 meters. Weight is 5800 kg.

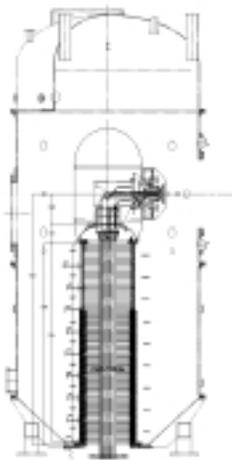

Figure 1 - Injector Assembly Column and Cathode Assembly installed in Vacuum Vessel

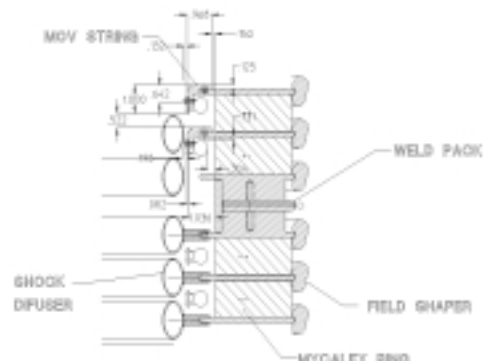

Figure 2 - Cross Section of Mycalex Rings

*Work supported by the US Department of Energy under contract DE-AC03-76SF00098

Column orientation was chosen to be vertical. This avoids high bending stress that would occur if the column were cantilevered horizontally.

Mycalex was the material chosen for the column insulating rings. Mycalex is a moldable and machinable ceramic material first developed about 1920. It is a matrix of mica flakes suspended in a glass matrix. It is dimensionally stable, has high thermal shock resistance, low outgassing rate, high resistance to tracking due to arc-overs. It has a notch test toughness of about 1.5 ft-lbs/in. as compared to ceramic at 0.5 ft-lbs/in. It's superior resistance over $Al_2O_3$ to cracking under arc-over conditions has been observed during the use of both materials in the high repetition rate induction machine at SRL, MA.[1]

During the design of the column, it was learned that there were mycalex rings of appropriate size (41.5" I.D. x 47.5" O.D. x 1.3" thick) existing in a decommissioned Van de Graaff accelerator at LANL. These rings proved to be almost ideal in all dimensions for use in the Darht-II Injector Column. Seventy rings were found that were spares and had never been installed in the Van de Graaff machine. Better yet, they were free. Unfortunately, there were only about 2/3 the quantity of mycalex rings needed for the column. After consideration of all options (new rings couldn't be made without costly tooling), it was decided to make up the ring short fall with $Al_2O_3$ rings.

Total hermiticity is required of all joints in the column in order to insure insulating oil does not migrate into the vacuum region and poison the source and contaminate the column surface. O-rings were not good enough. All brazing vendors we spoke with either no-bid or required an extensive development contract in order to develop the braze process required for this size work. Epoxy bonded joints were used entirely to make all ceramic to metal joints. Extensive testing of sample joints showed a tensile strength of 1400 psi normal and 5000 psi with the mycalex grain. Ceramic to metal joints exceeded 6000 psi tensile strength. All production joints were proof tested to 50,000 lbs. load. Hermiticity of all stack assemblies was measured with no helium leakage detected.

Tie rods along the inside diameter of the column were added to provide longitudinal compression of the column to counteract the 8700 kg force differential across the top end (vacuum "pressure") and to provide precompression for handling the column during handling and installation. Eight tie rods are used and produce a total of 24,000 kg compression.

Toroidal Field Shapers were attached to the outside diameter of the metal spacer rings to reduce the electric fields in the triple point regions to 4 kV/cm. The non-symmetrical shape is necessary to reduce the stress on the shaper lower surface to below 60 kV/cm. In order to reduce manufacturing cost and eliminate field stress risers due to joints and fasteners, the shapers were made in one piece. The shapers were assembled over the spacer rings by expanding them 4 mm by heating them to 200°C. When the shapers cool their inside groove engages the edge of the spacer rings. This process is shown in figure 3.

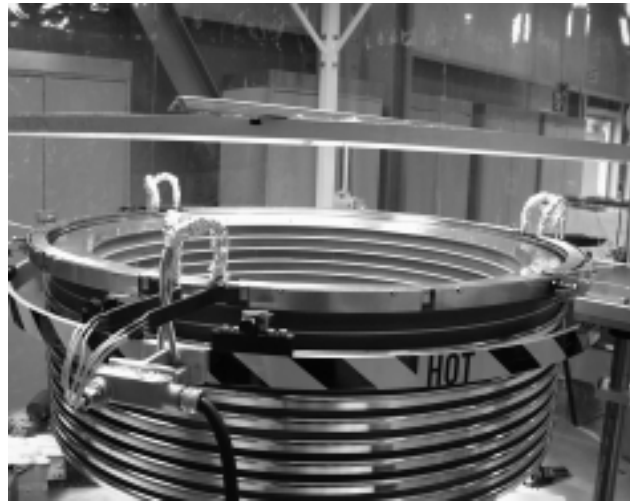

Figure 3 - Installation of Field Shapers

Oval shaped Shock Diffuser rings were attached to the inside edge of the spacer rings for the purpose of diffusing shock waves created by arcs in the oil. Each diffuser ring is supported with 12 radial fingers, which are springy in the radial direction so as not to transmit forces to the spacer/ceramic joint in the event of an arc to the ring. This arrangement is shown in figure 4. Also shown in this figure are the MOV and resistor board assemblies mounted between each spacer ring on the 6 lower stack assemblies.

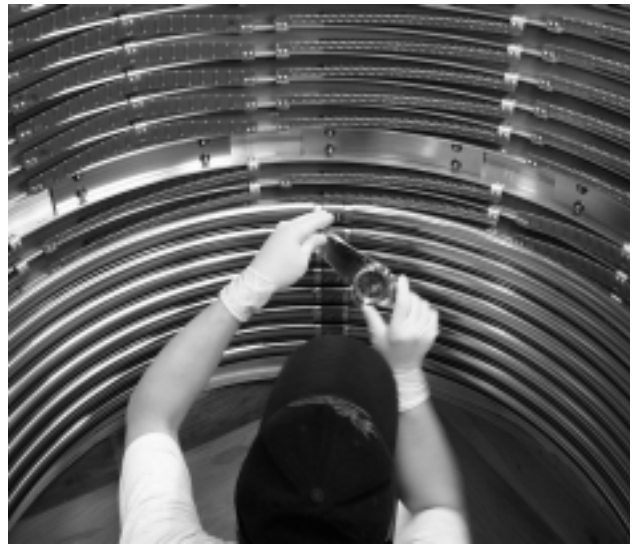

Figure 4 - Arrangement of MoV and Resistor Boards Installed Inside Stack Assembly

## 4 TESTING OF MATERIALS AND SCALE MODELS

Thirty 1" diameter x 1.3" long mycalex samples were cored out of spare mycalex rings. These samples were bonded to aluminum end pieces and were pull tested to failure. No necking of the samples was done. In all cases failure occurred near a bonded end. The epoxy never failed, as there was never any epoxy remaining on the mycalex part. The test determines the effective strength of the joint expected in the final column joints. The average strength was 1400 psi with a standard deviation of 100 psi.

Each mycalex ring used in the column was proof tested under internal hydrostatic pressure to test for rings with damage or internal flaws. These rings were old and their history and precise material was not well known. All rings were internally pressurized to 550 psi, which produces 3800 psi tensile stress (2/3 ultimate strength) in the hoop direction of the ring. Several failed and were discarded.

A ¼ scale stack was built and tested for shock survival. It was filled with oil and arcs were induced between electrodes in the oil to create shock waves. The results of these tests were analyzed and prediction made as to the stress and conservatism of the full-scale column. Bill Anderson, LANL, completed this analysis. The results indicate a safety factor of 4 to 10 depending on assumptions of the model. Each production stack assembly was vacuum leak checked below $10^{-9}$ std. cc/sec. Each production stack assembly was "Pull" tested to 50,000 lbs. tensile load to insure no cracks existed in the mycalex bond joints after assembly. One joint did fail at 43,000 lbs. and was repaired.

Electrical tests included "Hockey Puck" testing, which was to determine breakdown field stress levels and surface tracking behavior for the field shaper and bonded mycalex ring configuration. Three inch diameter mycalex disks (cut out of discarded 48" diameter rings) were bonded to stainless steel rings to form the electrically equivalent cross section of a production cross section. These tests indicate that we had a safety margin of at least a factor of 2 in voltage holdoff in our final design.

Stack Acceptance tests were completed on each of the 9 production stack assemblies. The fully configured stacks (including MOV and resistor strings) were operated under vacuum at high voltage to QA the electrical assembly work and determine final conditioning levels and operating margin for the final column assembly. The high voltage breakdown limit of the stack was tested to a factor of 2 above the operating voltage, and the energy and current through the MOV's and resistors were tested to greater than a factor of 1.5 above operating conditions.

## 5 SHOCK ANALYSIS

A detailed shock analysis of the column using the explicit non-linear finite element analysis program DYNA2D was completed by Joseph Stoner from LLNL. The general energy load assumption is based on information provided by Bill Anderson from LANL that 1.0 kJ of energy is deposited in a cylinder of oil extending from the "stalk" to the diffuser rings on the I.D. of the column. This initiates a hydrodynamic pressure pulse that travels through the oil impacting the Mycalex structure. A vertical compressive pressure was applied to the top of the Mycalex to represent the preload of the tie rod on the assembled stack.

The results from this 2½ dimensional axisymmetric analysis show that most of the energy in a shockwave produced by an arc will travel parallel to the axis of the stalk. Peak pressures of over 4000 psi will arrive at the mycalex wall approximately 45 microseconds after the arcing event. The maximum axial tensile stress is in the mycalex and is the stress of most concern; it is approximately 900 psi and occurs at a time of 60 microseconds.

The results of this analysis are not especially reassuring, as the factor of safety in the mycalex appears to be 1400/900 or 1.56. However, the results of this analysis are expected to be quite conservative and the stresses in the real event should be less than the values predicted from this simulation. Three arguments can be made to the conservative nature of the results: first, no damping has been modeled; second, no failure model has been used; third, the higher strength mechanical properties that usually accompany high strain rate dynamic shocks are not been modeled. An additional factor in favor of conservatism is that the volume of mycalex material subjected to high stress is extremely small. It is just the corner of the mycalex where it bonds to the aluminum spacer ring and azimuthally located only in the immediate vicinity of the line on the arc.

The best factor of safety for the mycalex is a combination of the analytical factor (1.56 x conservatism factors) above and the empirical factor (4 to 10) determined from the ¼ scale model experiment described above.